\documentclass{emulateapj}
\usepackage[english]{babel}
\usepackage{amsmath}
\usepackage{float}
\usepackage{natbib}
\usepackage[colorlinks=true,citecolor=blue,
  urlcolor=blue,breaklinks=true]{hyperref}
\usepackage{breakurl}
\usepackage{multirow}
\newcommand{\dd}{\mathrm{d}}
\newcommand{\minitab}[2][l]{\begin{tabular}{#1}#2\end{tabular}}

\begin{document}

\title{Predicting the Arrival Time of Coronal Mass Ejections with \\the Graduated Cylindrical Shell and Drag Force Model}
\shorttitle{Predicting the Arrival Time of CMEs}
\shortauthors{Shi et al.}
\author{Tong Shi$^{1,2}$, Yikang Wang$^{1,2}$, Linfeng Wan$^{1,2}$, Xin Cheng$^{1,2}$, Mingde Ding$^{1,2}$ and Jie Zhang$^3$}
\affil{$^1$School of Astronomy and Space Science, Nanjing University, Nanjing 210093, China; {\color{blue} dmd@nju.edu.cn}}
\affil{$^2$Key Laboratory for Modern Astronomy and Astrophysics (Nanjing University), Ministry of Education, Nanjing 210093, China}
\affil{$^3$School of Physics, Astronomy and Computational Sciences, George Mason University, Fairfax, VA 22030, USA}

\begin{abstract}
  Accurately predicting the arrival of coronal mass ejections (CMEs) at the Earth based on remote images is of critical significance in the study of space weather. In this paper, we make a statistical study of 21 Earth directed CMEs, exploring in particular the relationship between CME initial speeds and transit times. The initial speed of a CME is obtained by fitting the CME with the Graduated Cylindrical Shell model and is thus free of projection effects. We then use the drag force model to fit results of the transit time versus the initial speed. By adopting different drag regimes, i.e., the viscous, aerodynamics, and hybrid regimes, we get similar results, with the least mean estimation error of the hybrid model of 12.9 hours. CMEs with a propagation angle (the angle between the propagation direction and the Sun-Earth line) larger than its half angular width arrive at the Earth with an angular deviation caused by factors other than the radial solar wind drag. The drag force model cannot be well applied to such events. If we exclude these events in the sample, the prediction accuracy can be improved, i.e., the estimation error reduces to 6.8 hours. This work suggests that it is viable to predict the arrival time of CMEs at the Earth based on the initial parameters with a fairly good accuracy. Thus, it provides a method of space weather forecast of 1--5 days following the occurrence of CMEs.
\end{abstract}

\section{INTRODUCTION}

A coronal mass ejection (CME) is a massive eruption of plasma threaded with magnetic field from the Sun. A typical CME expels $10^{14-16}$ g of plasma and $10^{29-31}$ erg of kinetic energy \citep{1985JGR....90.8173H} through the solar corona into interplanetary space. In some extreme cases, CMEs have speeds over 2000 km s$^{-1}$ \citep{2004JGRA..109.7105Y, 2014NatCo...5E3481L} and may reach Earth within a day if they are Earth directed. CMEs are known to be the major source of interplanetary disturbances and are capable of influencing geomagnetic environments with shock waves, ejecta, and/or magnetic clouds \citep[cf.][]{1991JGR....96.7831G, 1993JGR....9818937G, 1994SoPh..153...73W, 1998JGR...103..277B, 2006SSRv..124..169K, 2007JGRA..11210102Z}. Some rare events with a very strong ejecta magnetic field, say, $Dst<-500$ nT, are potential hazards to spacecraft and some modern infrastructures \citep[e.g.,][]{1859MNRAS..20...13C, 2004SoPh..224..407C, 2006AdSpR..38..173S}. Therefore, understanding the CME propagation and ultimately making an accurate forecast of the CME arrival time are important issues in solar and space physics.

Since CMEs arrive at Earth in 1--5 days after they launch from the Sun \citep[e.g.,][]{1998GeoRL..25.3019B, 2000GeoRL..27..145G}, forecasting their arrival time with an accuracy of a dozen of hours is possible if we well understand the key factors controlling the CME propagation. Usually, studying the propagation process of CMEs requires remote sensing images taken during the initial stages of CMEs and in situ observations when CMEs reach the Earth. However, with only the observations of the initial stages at hand, some simple models are usually adopted to study the follow-up propagation process of CMEs. \citet{2000GeoRL..27..145G} developed a constant acceleration or deceleration model to account for the findings that fast CMEs experience a deceleration and slow ones tend to converge to the speed of the solar wind. \citet{2001JGR...10629207G} further improved the model in which the acceleration ceases before 1 AU accounting for the fact that slow CMEs have approximately the same transit time. This reduces the prediction error to 10.7 hours. More sophisticated approaches to CME kinematics consider the equation of motion governed by the drag force of the solar wind \citep{2001SoPh..202..173V, 2002JGRA..107.1019V, 2009A&A...498..885B, 2010A&A...512A..43V}. However, as can be seen in several validation studies \citep[e.g.,][]{2004AnGeo..22.4397O, 2013JGRA..118.6866C, 2014ApJS..213...21V}, the kinematical model yields a prediction error of around 10 hours based on the data set available, which is not significantly improved compared to the constant acceleration/deceleration model. Thus variants of the drag force model with its parameters determined based on different aspects have been further explored \citep{2004A&A...423..717V, 2010NatCo...1E..74B, 2014ApJ...792...49H, 2014ApJ...787..119M}.

An important topic in space weather forecasting is the estimation of the speed profile of the CME during its propagation. In particular, for the kinematical model, the CME initial speed is of crucial importance. The speed of a CME during the initial process can either be measured directly from the CME fronts in coronal images, or be derived by fitting the images taken from multi-perspectives such as the twin spacecraft STEREO A and B. The former method gives only the speed projected on the plane of sky. The latter, however, yields the true speed by using geometric triangulation techniques \citep{2010ApJ...710L..82L, 2010ApJ...722.1762L} and assuming that the leading edge in the images by different spacecraft corresponds to the same point. In the case of only one observer, some other geometric methods, e.g., the fixed-$\Phi$ fitting (FPF) \citep{1999JGR...10424739S, 2008GeoRL..3510110R}, the harmonic mean fitting (HMF) \citep{2010SoPh..267..411L, 2011ApJ...741...34M}, and the self-similar expansion fitting (SSEF) \citep{2012ApJ...750...23D, 2013SoPh..285..411M} have been adopted by further assuming a shape of the CME front. However, such geometric modelings consider CME propagation only in the ecliptic plane that may incur large errors in estimating the CME initial speed. The cone model \citep{1984ApJ...280..428F} and the Graduated Cylindrical Shell (GCS) model \citep{2006ApJ...652..763T, 2009SoPh..256..111T, 2011ApJS..194...33T} have also been used for a better estimation of CME speeds assuming a particular shape and self similarity of the CME. Although various models have been applied to case studies, previous statistical studies of a sample of CMEs were mostly restricted to the projection speeds with the drag model \citep[e.g.,][]{2000GeoRL..27..145G, 2001JGR...10629207G, 2004AnGeo..22.4397O}. It is not accurate enough, either, to assume a radial propagation from the source region since the CME propagation trajectory sometimes deviates from the radial direction \citep[e.g.,][]{2014ApJS..213...21V}.

In this paper, we present a statistical study of 21 Earth directed events. These events cause geomagnetic disturbances and serve as a good sample to test the prediction method of CME arrival time at the Earth. The GCS model is applied to determine the CME initial speeds in a three-dimensional perspective. A fitting on the results of the transit time versus the initial speed with the drag force model shows a mean absolute error of 12.9 hours. In particular, 5 CMEs are identified for their angular deflections due to mechanisms other than the solar wind drag, such as interactions with the background magnetic field or with other CMEs. The exclusion of these events yields a much better prediction accuracy of about 6.8 hours.

\section{OBSERVATIONS AND DATA ANALYSIS}

\subsection{The CME Sample}

CMEs have been observed for several decades but only in recent years is it possible to record the images of CME eruption simultaneously from different perspectives, such as those by Solar and Heliospheric Observatory (SOHO; \citealp{1995SoPh..162....1D}) and Solar Terrestrial Relations Observatory (STEREO; \citealp{2008SSRv..136....5K}). Among all the CMEs, the Earth directed ones are of particular interest since they are potential causes of geomagnetic storms. An Earth directed CME can be readily identified through the full or half halo shape (halo CME; \citealp{1982ApJ...263L.101H}) in the coronagraph of SOHO. When a CME propagates into the interplanetary space, it is termed as an ICME and can probably be observed by in situ instruments to record the parameters like the plasma density, speed, and magnetic field by spacecraft Global Geospace Science WIND \citep{1995SSRv...71....5A} and Advanced Composition Explorer (ACE; \citealp{1998SSRv...86....1S}) at 1 AU when the CME approaches the Earth.

The twin spacecraft STEREO-A/B are located separately and away from the Sun-Earth line, with A in an orbit ahead of the Earth and B behind the Earth. On board STEREO, the Sun Earth Connection Coronal and Heliospheric Investigation (SECCHI; \citealp{2008SSRv..136...67H}) contains two coronagraphs of COR1 and COR2 that continuously take white light images in the range of $1.5-4R_\odot$ and $2.5-15R_\odot$ and with a time resolution of 10 and 20 minutes, respectively. Along with the C2 and C3 instruments of Large Angle Spectral Coronagraphs (LASCO, \citealp{1995SoPh..162..357B}) on board SOHO that take white light images of a field of view (FOV) up to $32R_\odot$, the three coronagraphs provide a stereoscopic view of the CME and can track the CME front up to $15R_\odot$.

The data from the Heliospheric Imager (HI) on board STEREO is also used for verification of the interplanetary track of the CME. HI consists of two imagers. The FOV of HI1 is $20^\circ\times20^\circ$ centered at $14^\circ$ elongation from the center of the Sun and HI2 has a $70^\circ\times70^\circ$ FOV centered at $53^\circ.7$ from the center of the Sun. Thus, the instruments offer the possibility to track CMEs from near the Sun up to 1AU. Here, we employ J map techniques \citep{1999JGR...10424739S} to track the CME propagation and also verify whether a CME undergoes interactions with other CMEs.

In order to make a statistical study of the CME propagation times and test the drag force model, we employ the GMU CME/ICME list compiled by Phillip Hess and Jie Zhang (\url{http://solar.gmu.edu/heliophysics/index.php/GMU_CME/ICME_List}). We select the events with unambiguous shock fronts in the running difference images of COR2 and C2/C3. Our sample comprises of 21 events observed during 2008--2012 that are listed in Table \ref{cmelist}. In the following, the 2012 October 5 event is used to demonstrate our data processing.

\begin{deluxetable*}{ccccccccc}
\tablecolumns{9}
\tablewidth{0cm}
\tablecaption{List of the CME events studied in this work\label{cmelist}}
\tablehead{
\colhead{\#} & \colhead{onset time\tablenotemark{a}} & \colhead{$v_\text{init}$\tablenotemark{b} (km s$^{-1}$)} & \colhead{lon\tablenotemark{c} (${}^\circ$)} & \colhead{lat\tablenotemark{d} (${}^\circ$)} & \colhead{$\varphi\tablenotemark{e} ({}^\circ)$} & \colhead{ICME start\tablenotemark{f}} & \colhead{type\tablenotemark{g}} & \colhead{transit time (hr)}
}
\startdata
    1 & 2008-12-12 08:37 & $363 \pm 23$ & $6$ & $7$ & $20$ & 2008-12-17 02:00 & EJ & $113.4$ \\
    2 & 2010-04-03 09:54 & $864 \pm 7$ & $6$ & $-25$ & $38$ & 2010-04-05 08:00 & SH+MC & $46.1$ \\
    3 & 2010-04-08 03:39 & $512 \pm 34$ & $-16$ & $0$ & $28$ & 2010-04-11 12:00 & SH+EJ & $80.3$ \\
    4 & 2010-06-16 14:39 & $222 \pm 2$ & $-21$ & $2$ & $25$ & 2010-06-20 20:00 & EJ & $101.3$ \\
    5 & 2010-12-23 05:39 & $287 \pm 9$ & $18$ & $-22$ & $22$ & 2010-12-28 04:00 & EJ & $118.3$ \\
    6 & 2011-02-15 02:24 & $769 \pm 12$ & $6$ & $-11$ & $50$ & 2011-02-18 03:00 & MC+CIR & $72.6$ \\
    7 & 2011-03-25 05:39 & $90 \pm 3$ & $-34$ & $0$ & $21$ & 2011-03-29 16:00 & SH+EJ & $106.3$ \\
    8 & 2011-08-04 04:39 & $1512 \pm 90$ & $26$ & $21$ & $67$ & 2011-08-05 19:00 & SH & $38.4$ \\
    9 & 2011-09-06 22:54 & $678 \pm 13$ & $2$ & $13$ & $75$ & 2011-09-09 15:00 & SH+EJ & $64.1$ \\
    10 & 2011-09-14 00:39 & $505 \pm 5$ & $15$ & $19$ & $39$ & 2011-09-17 02:00 & SH+EJ & $73.3$ \\
    11 & 2011-10-22 10:39 & $882 \pm 4$ & $68$ & $57$ & $90$ & 2011-10-24 18:00 & SH+MC & $55.4$ \\
    12 & 2011-10-27 12:39 & $795 \pm 52$ & $-36$ & $25$ & $32$ & 2011-11-01 08:00 & SH & $115.3$ \\
    13 & 2012-01-19 14:54 & $1299 \pm 16$ & $-27$ & $42$ & $66$ & 2012-01-22 05:00 & SH & $62.1$ \\
    14 & 2012-03-05 03:54 & $1237 \pm 50$ & $-56$ & $29$ & $63$ & 2012-03-08 11:00 & SH+EJ & $79.1$ \\
    15 & 2012-03-13 17:39 & $1616 \pm 17$ & $51$ & $18$ & $78$ & 2012-03-15 13:00 & SH+EJ & $43.4$ \\
    16 & 2012-03-30 15:24 & $654 \pm 8$ & $-56$ & $30$ & $31$ & 2012-04-04 22:00 & EJ & $126.6$ \\
    17 & 2012-04-19 15:39 & $607 \pm 15$ & $-24$ & $-30$ & $50$ & 2012-04-23 02:30 & EJ+CIR & $82.8$ \\
    18 & 2012-07-12 16:39 & $1224 \pm 14$ & $0$ & $-11$ & $56$ & 2012-07-14 17:00 & SH+MC & $48.4$ \\
    19 & 2012-09-28 00:39 & $1104 \pm 112$ & $29$ & $10$ & $57$ & 2012-09-30 23:00 & SH+EJ & $70.3$ \\
    20 & 2012-10-05 03:39 & $558 \pm 21$ & $9$ & $-20$ & $35$ & 2012-10-08 05:00 & SH+MC & $73.3$ \\
    21 & 2012-10-27 16:54 & $340 \pm 28$ & $12$ & $10$ & $37$ & 2012-10-31 15:00 & SH+MC & $94.1$ \
\enddata
\tablenotetext{a}{{Date and time(UT) of the first appearance in the COR2 FOV.}}
\tablenotetext{b}{{Initial speed.}}
\tablenotetext{c}{{Longitude of the propagation direction fitted by the GCS model. Earth is at $0^\circ$ longitude; angles $>0^\circ$ corresponds to solar west.}}
\tablenotetext{d}{{Latitude.}}
\tablenotetext{e}{{Equivalent half angular width of the CME (see Section \ref{gcsfitdemo} and \ref{filter} for details).}}
\tablenotetext{f}{{Arrival date and time(UT) of the associated ICME.}}
\tablenotetext{g}{{ICME properties. CIR: co-rotating interaction regions; EJ: ejecta; MC: magnetic cloud; SH: shock.}}
\end{deluxetable*}

\subsection{GCS Model and the CME Speed}\label{gcsfitdemo}

The initial speed of a CME is determined based on the GCS model. First, the height of the CME apex (i.e., the shock front) is fitted with raytrace programs. As is shown below, the GCS model provides a good morphological approximation to the CME and a relatively robust height measurement.

\begin{figure}
  \plotone{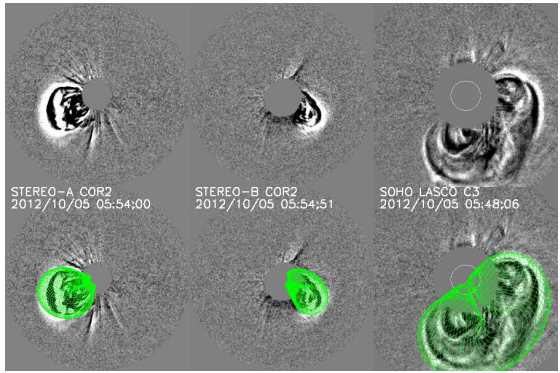}
  \caption{Running difference images of the 2012 October 5 CME at 05:54 UT. In the bottom row, the fitted GCS model is overlaid as the green wire frame.}
  \label{wireframe}
\end{figure}

For a clear identification of the CME morphology, we use the running difference images from the three coronagraphs on STEREO-A/B and SOHO (Figure \ref{wireframe}). We determine in total six parameters of CME including the longitude, latitude, tilt angle, half angular width, aspect ratio, and apex height that can best approximate the shape of the CME front. Except for the height, we assume that all the other five parameters do not change during the propagation of the CME, that is, the CME undergoes a self similar expansion into the space \citep{2012ApJ...750...23D, 2013SoPh..285..411M}. The goodness of fit under a certain set of parameters is reflected by the similarity between the wire-frame fitted CME shape and the actual shape shown in observed images. For each event, we finally choose the best set of parameters that can ensure that the goodness of fit for the time sequence of the event is most favorable.

The best fit for the 2012 October 5 event (see Figure \ref{wireframe} and Table \ref{gcspara}) gives the CME height of 10.72 $R_\odot$ at 05:54 UT. Note that an automatic similarity test is conducted after the set of parameters is derived manually. The test examines the difference between the shape of the fitted wire frame and the observed CME profile shown in the images of STEREO-A/B. For this event, the similarity reaches a level of 68.7\%. The error of the six parameters is estimated by varying the corresponding parameter that results in a 10\% change in similarity. The uncertainty of the height is of an order less than $1 R_\odot$. Compared to the relatively large FOV of COR2 (15$R_\odot$), the uncertainty in height is small. Thus, the height is a robust parameter in the GCS model fitting even though the other five parameters may involve relatively large uncertainties.

\begin{deluxetable*}{cccccc}
\tablewidth{0cm}
\tablecaption{Best fitted parameters of the 2012 October 5 CME by the GCS model\label{gcspara}}
\tablehead{
\colhead{lon\tablenotemark{a} ($^\circ$)} & \colhead{lat\tablenotemark{b} ($^\circ$)} & \colhead{$\Gamma$\tablenotemark{c} ($^\circ$)} & \colhead{$\alpha$\tablenotemark{d} ($^\circ$)} & \colhead{$\kappa$\tablenotemark{e}} & \colhead{H\tablenotemark{f} ($R_\odot$)}
}
\startdata
$54.4_{-7.0}^{+16}$ & $-17.3_{-7.0}^{+3.4}$ & $39.1_{-23}^{+22}$ & $27.4_{-15}^{+19}$ & $0.410_{-0.18}^{+0.11}$ & $10.72_{-0.55}^{+0.80}$
\enddata
\tablenotetext{a}{{Carrington longitude.}}
\tablenotetext{b}{{Latitude.}}
\tablenotetext{c}{{Tilt angle respect to the equator, with counterclockwise being the positive.}}
\tablenotetext{d}{{Half angular width between the two legs of the model.}}
\tablenotetext{e}{{Aspect ratio.}}
\tablenotetext{f}{{CME apex height.}}
\end{deluxetable*}

\begin{figure}
  \plotone{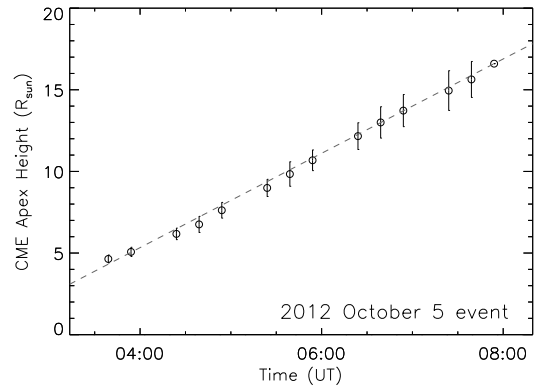}
  \caption{Height versus time curve of the 2012 October 5 event. The height refers to that of the CME apex obtained by the GCS fitting. The error bar indicates the standard deviation from 6 independent measurements. The initial speed $v_\text{init}=558\pm 21$ km s$^{-1}$ is obtained through a linear fit to the curve.}
  \label{hvt}
\end{figure}

A good fit yields a height-time relationship for the CME propagating into the space. In practice, we repeat the GCS fitting 6 times for each event and then make an average of the fitted parameters. The height-time curve for the 2012 October 5 event is shown in Figure \ref{hvt}. The error bars represent the standard deviation of the 6 measurements. The initial speed of the CME is then obtained through a linear fit to the curve assuming that the CME has already reached a static speed in the FOV of COR2 \citep{2006ApJ...649.1100Z}. For this event, $v_\text{init}$ is $558 \pm 21$ km s$^{-1}$. The error refers to the 3-$\sigma$ uncertainty of the linear fitting.

\subsection{Transit Time}

\begin{figure}
  \epsscale{1}
	\plotone{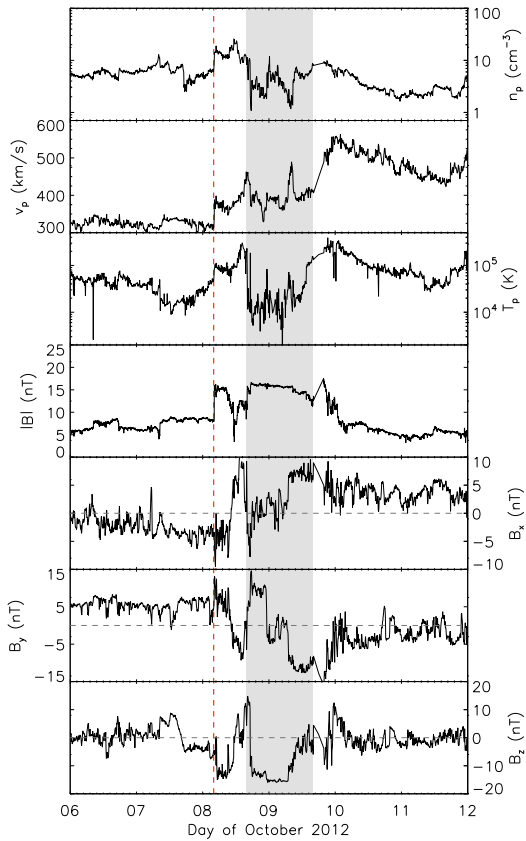}
	\caption{Plasma and magnetic field parameters of the ICME observed by WIND. The vertical dashed line shows the arrival of the shock while the shaded region indicates the magnetic cloud interval.}
	\label{insitu}
\end{figure}

The CME transit time is the time elapsed from the occurrence of the CME to its arrival at 1AU. This parameter can be directly obtained from the in situ data by WIND (see Figure \ref{insitu}). In the 2012 October 5 event, a clear shock is seen at 05:00 UT on October 8 as indicated by the sudden jumps in the proton density, speed, and temperature (as denoted by the vertical dashed line in the figure). Behind the shock is a sheath region with enhanced proton density and temperature and variable magnetic field. A Magnetic Cloud (MC; \citealp{1981JGR....86.6673B}) is then identified by a strong magnetic field, a smooth rotation of the field, and a depressed proton temperature (marked by the shaded region). The MC ends at 17:00 UT on 2012 October 9. A connection between the CME in remote-sensing images and the ICME observed in situ can be established with the HI data in the form of monthly movies or J maps. Thus, the transit time is calculated to be the time between the first appearance of the CME in the COR2 FOV and the detection of shock near the Earth.

\section{STATISTICAL RESULTS AND PREDICTION MODELS}

\begin{figure}
	\epsscale{1.2}
  \plotone{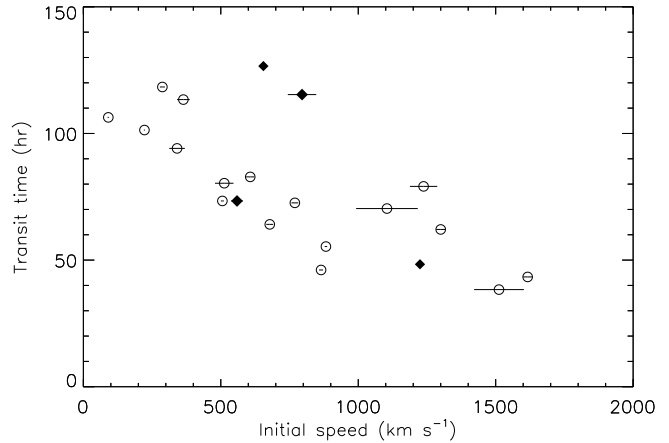}
  \caption{Transit time versus initial speed for 21 CME/ICME events. The horizontal bars denote the 3-$\sigma$ error from the linear fit. The four diamonds mark the events (the 2012 July 12, 2012 October 5, 2011 October 27, and 2012 March 30 events) that serve as examples for CME tracking in interplanetary space with J maps. The 2012 October 5 event is also used as an example for the GCS model fitting.}
  \label{TTvVinit}
\end{figure}

Using the methods described above, we derive the initial speed and the transit time for 21 CME events in our sample. Figure \ref{TTvVinit} plots the two parameters showing a close relationship between them. Note that the 2012 October 5 event and other three events serve as demonstrations in the following sections. As expected, the faster the CME's initial speed, the shorter time it takes for the CME to propagate to the Earth, which is consistent with previous observations \citep{2001JGR...10629207G}. There is also a fairly large scatter of the data points as shown in Figure \ref{TTvVinit}. Besides the measurement errors in the speed and transit time, the CME may encounter different forces (such as angular forces exerted by the background magnetic field and CME-CME interactions) as discussed below.

\subsection{The Drag Force Model}

In general, CMEs interact with the solar wind in three perspectives: the Lorentz force on the plasma and the threaded magnetic flux rope, the gravity, and the drag force \citep{1996JGR...10127499C}. Compared with the drag force, the other two can be neglected \citep{2002JGRA..107.1019V}, given the reduction of the magnetic field in the heliosphere \citep{1989ApJ...344.1051C, 1990SoPh..129..295V, 2001JGR...10625249V} and the low density of CMEs \citep{2001SoPh..202..173V}. Thus, the drag force by the solar wind is a dominant force controlling the propagation of CMEs in the interplanetary space.

By considering the one-dimensional problem, the equation of motion can be expressed as
\begin{equation}
\frac{\dd v}{\dd t} = -\gamma (v-w) |v-w|^{\beta-1}.
\label{eqmotion}
\end{equation}
The exponent $\beta$ describes the drag regime; usually $\beta$ can vary in the range of 1--2, with $\beta=1$ for viscous and $\beta=2$ for aerodynamics regime \citep{2010NatCo...1E..74B}.

The parameter $\gamma$ is expressed as $\gamma = \mathcal{C}_d A \rho_w / m$ \citep{2010A&A...512A..43V}, where $\mathcal{C}_d$ is the drag coefficient, $m$ is the mass, and $A$ is the cross section of the CME. We take $\mathcal{C}_d$ as a constant of order unity \citep{2004SoPh..221..135C}. Assuming that CMEs undergo a self-similar expansion into the space, the cross section can be calculated as $A \simeq \pi (r \varphi)^2 / 4$, where $\varphi$ is the half angle, and $r$ is the height of the center of the CME shell. The solar wind density can be given by the empirical formula $\rho_w \simeq \rho_1 / (r/R_\odot)^2$ \citep{1998SoPh..183..165L}, where $\rho_1$ is the empirical solar wind density at 1 AU. Therefore, $\gamma \approx \pi \mathcal{C}_d R_\odot^2 \rho_1 \varphi^2/4m \propto \varphi^2/m$. According to the GCS model, most of the CME mass is concentrated on the surface of the shell. Thus, $\varphi^2/m$ does not change with CME propagation and is proportional to the initial surface density of the CME, which, for simplicity, we regard as nearly the same for different CMEs. Therefore, the parameter $\gamma$ is a constant in our modeling. However, in some previous work \cite[e.g.,][]{2002JGRA..107.1019V, 2010NatCo...1E..74B}, the parameter $\gamma$ is taken to be inversely related to the CME height. This relation is found statistically from SOHO data within the FOV of LASCO C3 of 30 $R_\odot$ \cite{2001SoPh..202..173V}, so it may not work for CME propagation in the interplanetary space.

The equation of motion can then be integrated analytically to yield
\begin{equation}
\begin{aligned}
x_\beta = wt&\pm\frac{1}{\gamma(2-\beta)}\Large\{|v_\text{init}-w|^{2-\beta}- \\
& \left[\gamma(\beta-1)t+|v_\text{init}-w|^{1-\beta}\right]^{\frac{2-\beta}{1-\beta}}\Large\}\label{eqhybrid},
\end{aligned}
\end{equation}
where $1<\beta<2$ and the first positive sign is for $v_\text{init}>w$ and the first negative sign is for $v_\text{init}<w$. In particular, for the pure viscous ($\beta=1$) regime, we have
\begin{equation}
x_1=wt+\frac{1}{\gamma}(v_\text{init}-w)\left(1-e^{-\gamma t}\right),\label{eqvisc}
\end{equation}
and for the pure aerodynamic ($\beta=2$) regime, we get
\begin{equation}
x_2=wt \pm \frac{1}{\gamma}\ln\left(\gamma|v_\text{init}-w|t+1\right).\label{eqaero}
\end{equation}
In the follows, we refer Equation (\ref{eqhybrid}) to the hybrid (drag force) model, as discriminated from the viscous model by Equation (\ref{eqvisc}), and the aerodynamics model by Equation (\ref{eqaero}).

Given a certain set of parameters $\gamma$, $\beta$, and $w$, the equation of motion can be solved to yield the relationship between the initial speed and the transit time to 1 AU. Thus, the theoretical transit time can be expressed as $T=T(v_\text{init}; \gamma, \beta, w)$. We then get the parameters $\gamma$, $\beta$, and $w$ by fitting the measured initial speed versus transit time from the sample with the theoretical relationship.

\subsection{Fitting Results for the Whole Sample}

\begin{figure*}
	\plotone{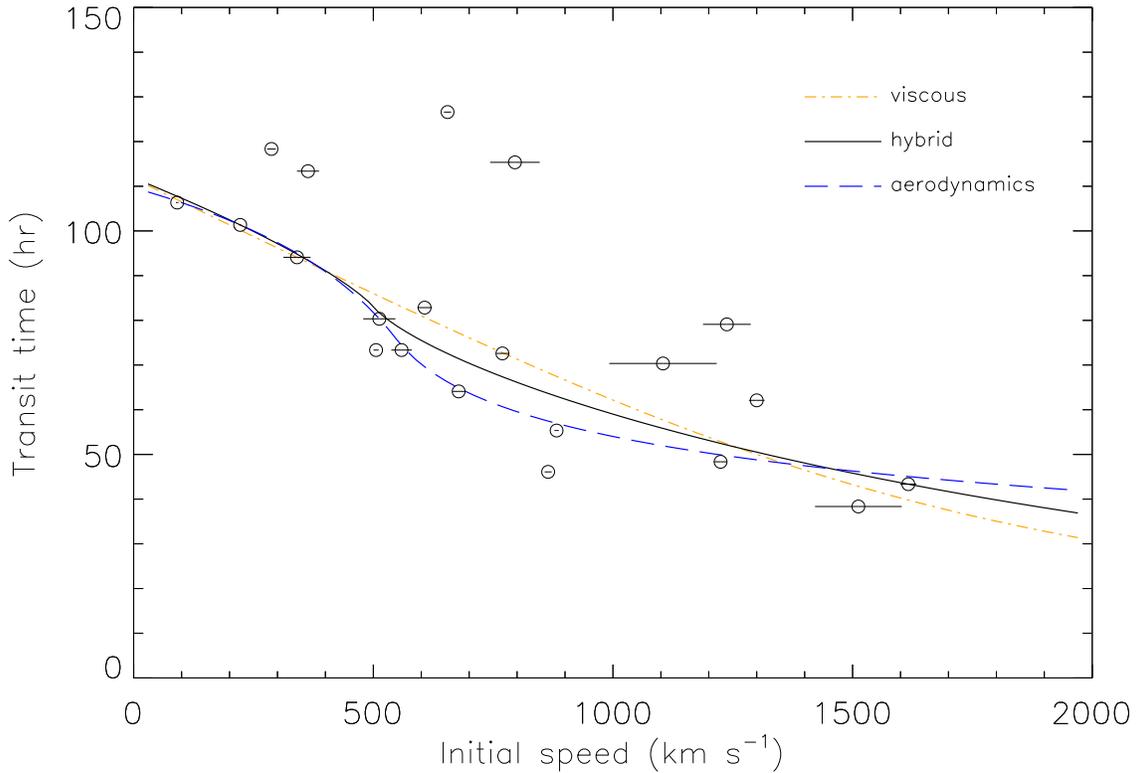}
	\caption{Fitting to the transit time-initial speed measurements for the whole sample by the drag force model. The solid black curve refers to the best fit by the hybrid model, with a mean absolute estimation error $\langle|\Delta\tau|\rangle=12.9$ hr. The dash-dotted orange curve is for the viscous model with $\langle|\Delta\tau|\rangle=13.2$ hr. The dashed blue curve is for the aerodynamics model with $\langle|\Delta\tau|\rangle=13.2 $ hr.}
	\label{tvs}
\end{figure*}

We first fit the transit time-initial speed distribution for the whole sample containing the 21 CME events using the drag force model, as shown in Figure \ref{tvs}. We implement the nonlinear least absolute curve fitting with the Levenberg-Marquardt algorithm \citep{2009ASPC..411..251M}. This technique allows faster convergence to the local minimum but dependents on the initial value of the parameters.  Considering the possible existence of several local minima in fitting errors, we choose the initial values from the parameter space to ensure that the final error is the smallest. However, after several test runs, we find that the mean absolute error in the prediction of transit time ($\langle|\Delta\tau|\rangle$) varies little (less than 0.1 hr) subject to the change of the parameters. Therefore, as explained above, despite the mathematical difficulty, the fitted parameters should result in a reasonable solution.

The parameters obtained from the fitting are listed in Table \ref{fitpara}. Fitting with the hybrid model gives a mean absolute error $\langle|\Delta\tau|\rangle = 12.9$ hr. Here $\Delta\tau$ is defined as the difference between the predicted transit time and the observed one; therefore, $\Delta\tau<0$ refers to cases in which the former is somewhat shorter than the latter. We also calculate the average value of the time difference, termed as the mean error $\langle\Delta\tau\rangle$, which represents a systematic deviation of the observed transit time from the theoretical one. It may be caused by some physical processes that cannot be described in terms of the drag force model. For the hybrid model, $\langle\Delta\tau\rangle=-7.1$ hr. The fitting by the pure viscous and aerodynamic models both give $\langle|\Delta\tau|\rangle =$ 13.2 hr, with $\langle\Delta\tau\rangle = -5.1$ hr and $-9.9$ hr, respectively.

\begin{deluxetable*}{c|ccccccc}
\tablewidth{0cm}
\tablecaption{Best fitted parameters of the drag force model\label{fitpara}}
\tablehead{\colhead{} & \colhead{} & \colhead{viscous} & \colhead{hybrid} & \colhead{aerodynamics}}
\startdata
    \multirow{5}*{\minitab[c]{Whole\\Sample\tablenotemark{a}}}
    & $\gamma$\tablenotemark{c} & $(1.12_{-0.37}^{+0.66})\times 10^{-5}$ & $(1.30_{-0.46}^{+0.84})\times 10^{-6}$ & $(3.02_{-1.17}^{+1.82})\times 10^{-8}$\\
    & $\beta$\tablenotemark{d} & $1$ & $1.37_{-0.07}^{+0.10}$ & $2$\\
    & $w$\tablenotemark{e} & $477_{-42}^{+48}$ & $501_{-67}^{+41}$ & $549_{-94}^{+36}$\\
    & $\langle|\Delta\tau|\rangle$\tablenotemark{f} & $13.2$ & $12.9$ & $13.2$\\
    & $\langle\Delta\tau\rangle$\tablenotemark{g} & $-5.1$ & $-7.1$ & $-9.9$\\
    \tableline
    \multirow{5}*{\minitab[c]{Restricted\\Sample\tablenotemark{b}}}
    & $\gamma$ & $(0.90_{-0.18}^{+0.39})\times 10^{-5}$ & $(1.10_{-0.27}^{+0.31})\times 10^{-7}$ & $(2.71_{-0.56}^{+0.67})\times 10^{-8}$\\
    & $\beta$ & $1$ & $1.76_{-0.05}^{+0.05}$ & $2$\\
    & $w$ & $488_{-25}^{+54}$ & $546_{-38}^{+26}$ & $558_{-31}^{+19}$\\
    & $\langle|\Delta\tau|\rangle$ & $8.0$ & $6.8$ & $6.7$\\
    & $\langle\Delta\tau\rangle$ & $-0.7$ & $-3.3$ & $-3.5$
\enddata
\tablenotetext{a}{Parameters obtained by fitting the whole sample including 21 CME events.}
\tablenotetext{b}{Parameters obtained by fitting the restricted sample including 16 CME events.}
\tablenotetext{c}{The coefficient proportional to the strength of the drag force.}
\tablenotetext{d}{The exponent on the difference of speed of the CME and the solar wind.}
\tablenotetext{e}{The solar wind speed.}
\tablenotetext{f}{The mean absolute error between the observed transit time and the predicted one.}
\tablenotetext{g}{The mean error representing the systematic under- or over-estimation.}
\end{deluxetable*}

\subsection{Fitting Results for a Restricted Sample}\label{filter}

The three-dimensional propagation of a CME is largely related to the ambient environment. A non-uniformly distributed coronal magnetic field may result in an asymmetric or non-radial motion of the CME in the early stage of propagation \citep[e.g.,][]{1986JGR....91...31M, 2000JASTP..62.1457G, 2009AnGeo..27.4491K, 2011SoPh..269..389S, 2013SoPh..287..391P}. A CME can also be deflected during its propagation in the interplanetary space when interacted with the non-radial magnetic backgrounds with the deflection angle dependent on the speed of the CMEs \citep[e.g.,][]{2004SoPh..222..329W, 2014JGRA..119.5117W}. The drag force model cannot be well applied to such CMEs with such events with non-radial forces. Here we denote by $\theta$ the propagation angle of a CME, which is the angle between the central axis of the CME and the Sun-Earth line, and by $\varphi$ the CME half angular width. In principle, a CME with $\varphi < \theta$ that propagates radially and self-similarly initially should not encounter the Earth; however, it might reach the Earth through angular deflections. Such events may worsen the fitting results and should be analyzed separately.

The half angle $\varphi$ of a CME is calculated based on the cone model instead of the GCS model in order to reduce the number of free parameters. Since the tilt angle $\Gamma$, half angle $\alpha$, and aspect ratio $\kappa$ in the GCS model are partially degenerated and each of them involves a large measurement error (see Table \ref{gcspara} for the error derived from a similarity test), we combine these three parameters into one, which just corresponds to the half angle in the cone model. From the definition of the GCS model \citep{2009SoPh..256..111T}, we may derive that
\begin{equation}
\varphi \simeq \left[\alpha (\alpha+\delta)\right]^{1/2},
\end{equation}
where $\kappa=\sin\delta$ and $\Gamma$ is eliminated.

\begin{figure*}
	\plotone{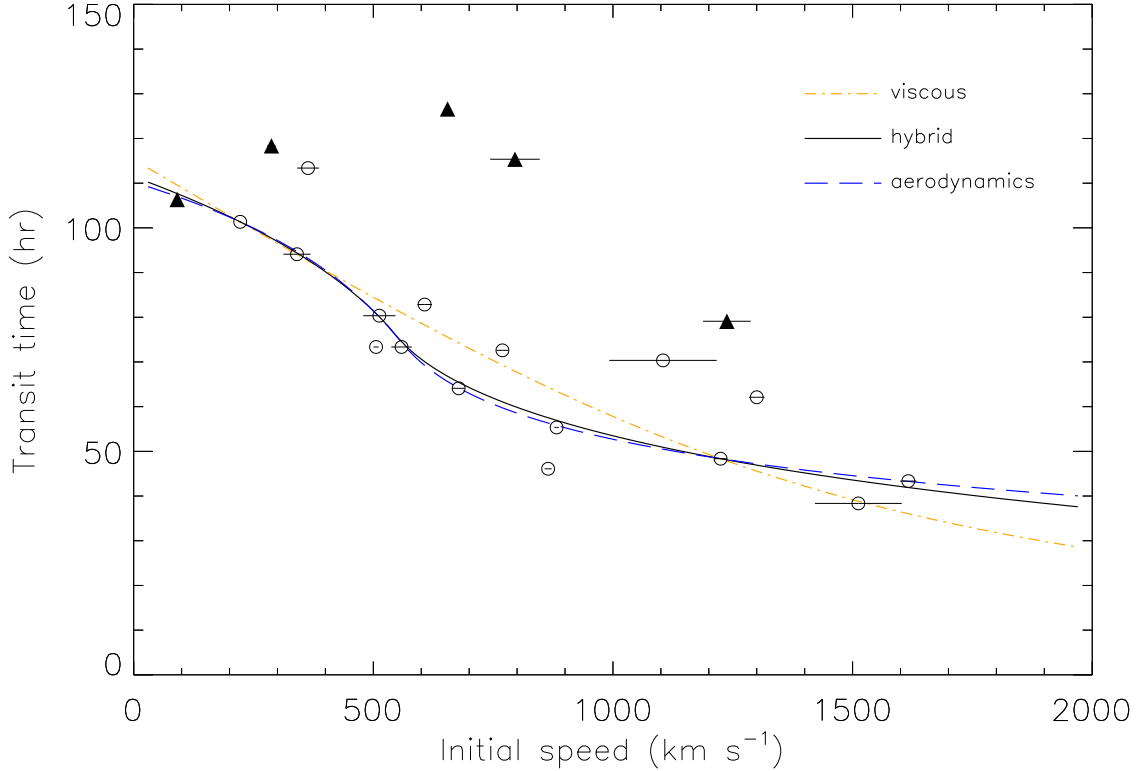}
	\caption{Fitting to the transit time-initial speed measurements for the restricted sample by the drag force model. The filled triangles refer to those events that are excluded from the sample (see text for details). The solid black curve refers to the best fit by the hybrid model, with a mean absolute estimation error $\langle|\Delta\tau|\rangle=6.8$ hr. The dash-dotted orange curve is for the viscous model with $\langle|\Delta\tau|\rangle=9.0$ hr. The dashed blue curve is for the aerodynamics model with $\langle|\Delta\tau|\rangle=6.7$ hr.}
	\label{filtered_fit}
\end{figure*}

As mentioned above, those events with the criteria $\varphi < \theta$ may undergo obvious interactions with non-radial forces that make their propagation direction deviate largely from radial. After a careful examination, we seek out 5 events that possibly belong to this category and exclude them in the sample of CMEs. We then redo the fitting using the viscous, hybrid, and aerodynamics model to the measured results of the remainder 16 events, as is shown in Figure \ref{filtered_fit}. Note that most of the filtered-out events (4 out of 5) deviate largely from the theoretical curve, implying that the drag force models do not apply to them. With the restricted sample, the estimation errors are 8.0, 6.8, and 6.7 hours for the viscous, hybrid, and aerodynamics model, respectively.

It is seen that the estimation error by the aerodynamics model and that by the viscous model are equally large when using the original sample. However, both of them are significantly improved if we use the restricted sample instead. Comparatively, the aerodynamics model yields an even smaller estimation error. Thus, the aerodynamics drag might play an important role in the solar wind drag in the interplanetary space. The hybrid model, however, fails to produce a more accurate estimation even with an extra free parameter than the aerodynamics model. This is because of the existence of a singularity near $\beta=2$ for the hybrid model (Equation \ref{eqhybrid}) where the calculated CME height diverges. Nevertheless, the systematic error of the hybrid model is smaller than that of the aerodynamics model. Recently, \citet{2010NatCo...1E..74B} obtained a fitted value of the $\beta$ parameter to be 2.27. Our fitting using the hybrid model gives a $\beta$ value of 1.76, which seems not sufficiently large to support the dominant role of the aerodynamics model. Such a difference may be due to our use of a different equation of motion without the radially decreasing term. Judging from our statistical results and the model fitting, the solar wind drag can be represented by a combination of the viscous and aerodynamics drags, with the latter likely playing a major role.

For the 5 events excluded here, the predicted transit time is less than their real transit time, suggesting that these CMEs travel longer distances and/or at slower speeds than what are predicted by the model. The real distances are elongated by angular deflections. Moreover, CME-CME interactions can slow down the CME propagation speed.

\subsection{Verifying the Interplanetary Propagation of CMEs}

\begin{figure*}
  \plotone{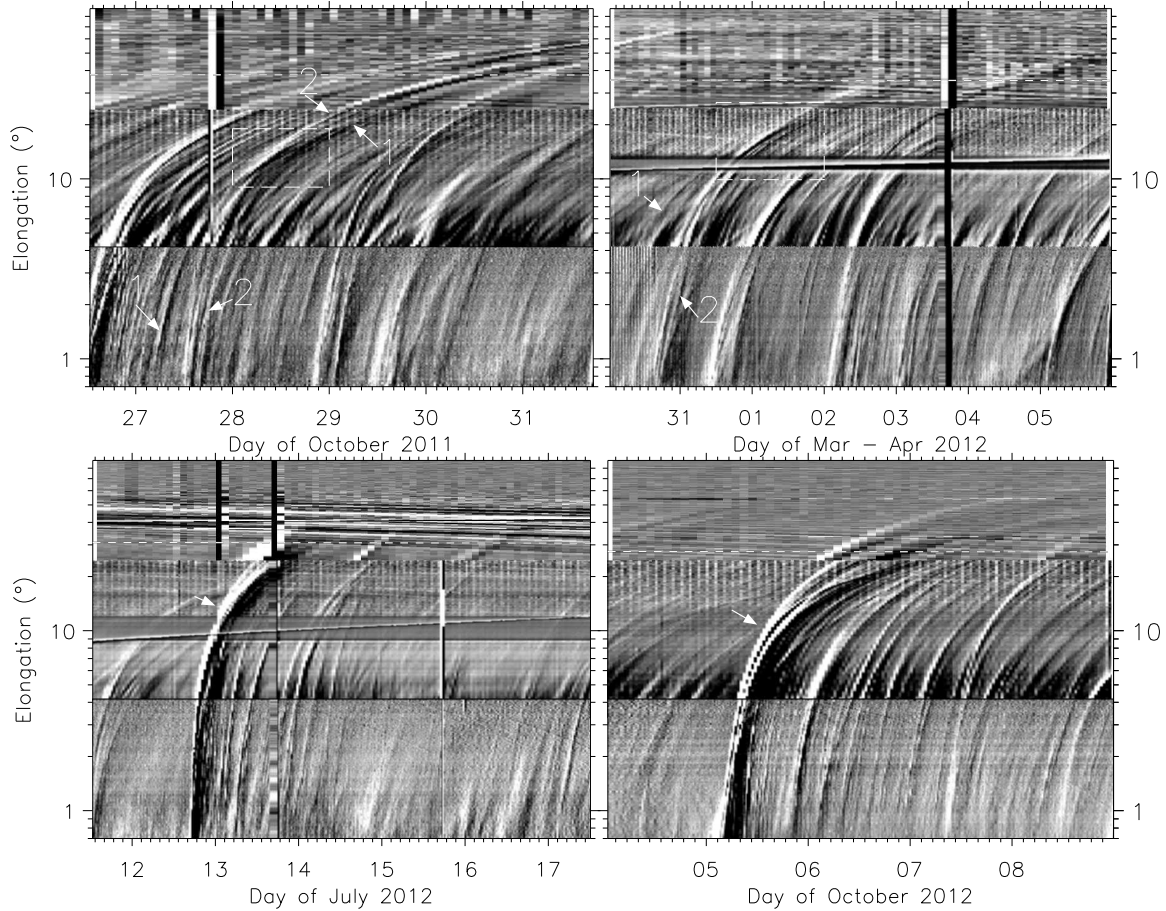}
  \caption{Time-elongation maps (J maps) constructed from COR2, HI1, HI2 images by STEREO A with a slice along the ecliptic plane. The contrast is rescaled for each row. Upper Left: The 2011 October 27 event. The two arrows point to two CME events in which CME 2 has a higher velocity. The small box indicates the location where the two CMEs may undergo an interaction. Upper Right: The 2011 March 30 event. The small box is the region where the two CMEs seem to merge. Lower Left: The 2012 July 12 event. Lower Right: The 2012 October 5 event. The latter two events show no indication of interplanetary interactions.}
  \label{jmaps}
\end{figure*}

We check in detail four representative events in our sample. The 2011 October 27 event has an observed transit time of 115.3 hours. The predicted transit time based on the 3 models, however, is less than 80 hours. Such a discrepancy suggests that the predicted propagation speed is obviously higher than the actual one. The J map, which is constructed from the running difference images along the ecliptic plane by STEREO-A, is shown in Figure \ref{jmaps}. The figure reveals two CMEs, indicated by the two arrows, the second of which is actually the event included in our sample, starting at 12:39 UT. One can obviously see that the faster one (noted as CME-2) catches up with the previous slower one (noted as CME-1) at a place (shown as the boxed region) where they undergo an interaction. Thus, the faster CME we are tracking should be exerted on an extra resistance force other than the solar wind drag. A similar process is also shown in the 2012 March 30 event. Another possible result of the interactions is the angular motion of the CMEs. Since in these two events the angular width of the CME is less than the propagation angle, interactions with a previous CME might cause the propagation direction deviating from radial. In contrast, the 2012 July 12 and the 2012 October 5 events do not show significant interplanetary interactions, so that the drag force model can be well applied to such cases.

\section{DISCUSSION AND CONCLUSION}

In this paper, we make statistical study of the transit time (between the Sun and the Earth) against the initial speed for a sample including 21 CME events directed towards the Earth. For each event, the in situ signal clearly shows at least one of the three identities (shock, ejecta, or magnetic cloud) for a typical ICME, ensuring that the CME does actually reach the Earth and disturb the geomagnetic environment. The initial speeds of these CMEs cover a range from 90 to 1616 km s$^{-1}$, which is sufficiently wide to cover both slow and fast ones.

The initial speed of an event is determined by the GCS model fitting the images obtained from COR2 on board STEREO A/B and C2/3 on board SOHO. The images from three viewing angles offer a much better accuracy in determination of the CME height and velocity, which are, in particular, free from the projection effect.

The drag force model is then applied to fit the transit time versus initial speed distribution. We achieve mean absolute errors of 13.2, 12.9, and 13.2 hours for the viscous, hybrid, and drag force model, respectively. These values are comparable with previous results \citep[e.g.,][]{2001JGR...10629207G, 2004AnGeo..22.4397O, 2013JGRA..118.6866C, 2014ApJS..213...21V}. However, since the drag force model considers only the radial drag force exerted by the solar wind, it cannot apply to those events that undergo interplanetary interactions with other CMEs and whose propagation direction deviates obviously from radial. We pick out 5 events that likely belong to such a category and exclude them in the statistical sample. Doing so greatly improves the prediction accuracy. The estimation error of the transit time is reduced to be 8.0, 6.8, and 6.7 hours for the three models, respectively, using the restricted sample. Moreover, the result shows that the higher order aerodynamic model achieves a better prediction than the pure viscous model. If using the hybrid model, the best fitted exponent is 1.76, suggesting that the solar wind drag is a combination of the two drag regimes with the aerodynamics drag somewhat in more effect. Since the average time for CME propagation in our sample is 79.3 hours, an error of 6.7 hours amounts only to about 8\%. Thus, it is practical to use the drag force model to predict the CME arrival time at the Earth with a fairly good accuracy. Nevertheless, to reach a higher accuracy, even more sophisticated models are needed to account for the events with non-radial interactions.

Recently, \citet{2014ApJS..213...21V} conducted the CME arrival time prediction based on the drag force model fitting and MHD simulation. They reported estimation errors of about 14 hours. Here, we use the hybrid model and achieve a similar estimation error of 12.9 hours for the whole data sample but a much improved error of 6.8 hours for the restricted sample. Besides the model, an accurate determination of the CME initial speeds is very crucial. By checking the events in our sample, we find that for some of them, using only the imaging observations from a single instrument, assuming a radial propagation from the located source region, may not yield accurate results because of the possible angular deflections of the CMEs at the initial stage in the lower corona. A stereoscopic determination of the CME morphology and propagation are thus needed to deduce the CME initial speed. This is the main reason why we can achieve a somewhat better prediction accuracy compared with previous studies. Our method is based on the observations in the earlier process of the CME propagation, which is within the FOV of COR2 (15 $R_\odot$). Thus, our method can be used to forecast the CME arrival time once the event is erupted. However, a better estimation accuracy may be achieved by tracking the event for an extended distance with HI \citep[e.g.,][]{2014ApJ...787..119M}.

The three-dimensional nature of the CME morphology also adds to the uncertainties in the transit time estimation. Since CMEs have a finite angular width, those with a propagation direction slightly deviated from the Sun-Earth line may still encounter the Earth at some off-axis point. In this case, the CME forehead may have traveled a distance larger than 1 AU. Therefore, a correction based on the angular width and propagation angle of the CME is needed \citep{2014JGRA..119.5107S}. For example, typical parameters of $\varphi=45^\circ$ and $\theta=10^\circ$ gives $\Delta t \approx 1.8$ hr, while $\varphi=60^\circ$ and $\theta=30^\circ$ results in $\Delta t \approx 15.5$ hr for the correction to the transit time.

Observations have also revealed that many CMEs have two different fronts of a bubble-shaped structure and a flux-rope-shaped structure \citep{2014ApJ...794..148K}. In our study, we only model the outermost front seen in the running difference images. Thus, a confusion between the two fronts, whose speeds differ, is possible and may also affect the accuracy of determination of the CME initial speed.

\acknowledgments
We are grateful to the referee for valuable comments. This work was supported by NSFC under grants 11303016, 11373023, and NKBRSF under grants 2011CB811402 and 2014CB744203.

\end{document}